\documentclass[pdflatex,sn-aps, iicol]{sn-jnl}

\usepackage{graphicx}%
\usepackage{multirow}%
\usepackage{amsmath,amssymb,amsfonts}%
\usepackage{amsthm}%
\usepackage{mathrsfs}%
\usepackage[title]{appendix}%
\usepackage{xcolor}%
\usepackage{textcomp}%
\usepackage{manyfoot}%
\usepackage{booktabs}%
\usepackage{algorithm}%
\usepackage{algorithmicx}%
\usepackage{algpseudocode}%
\usepackage{listings}%
\usepackage{braket}
\usepackage{natbib}
\usepackage{subcaption}
\usepackage{siunitx}
\usepackage{bm}
\usepackage{subcaption}
\theoremstyle{thmstyleone}%
\theoremstyle{thmstyletwo}%

\theoremstyle{thmstylethree}%

\begin{document}







\title[Article Title]{Long-lived quantum memories with rare-earth ions doped in crystals}

\author[1,2,3]{\fnm{Yu-Yu} \sur{Sa}}\email{syy200406@mail.ustc.edu.cn}

\author[1,2,3]{\fnm{Pei-Xi} \sur{Liu}}\email{liupeixi@mail.ustc.edu.cn}

\author*[1,2,3,4]{\fnm{Zong-Quan} \sur{Zhou}}\email{zq\_zhou@ustc.edu.cn}
\author[1,2,3,4]{\fnm{Chuan-Feng} \sur{Li}}
\affil[1]{\orgdiv{Laboratory of Quantum Information}, \orgname{University of Science and Technology of China}, \orgaddress{\city{Hefei}, \postcode{230026}, \state{Anhui}, \country{China}}}

\affil[2]{\orgdiv{Anhui Province Key Laboratory of Quantum Network}, \orgname{University of Science and Technology of China}, \orgaddress{\city{Hefei}, \postcode{230026}, \state{Anhui}, \country{China}}}

\affil[3]{\orgdiv{CAS Center For Excellence in Quantum Information and Quantum Physics}, \orgname{University of Science and Technology of China}, \orgaddress{\city{Hefei}, \postcode{230026}, \state{Anhui}, \country{China}}}

\affil[4]{\orgdiv{Hefei National Laboratory}, \orgname{University of Science and Technology of China}, \orgaddress{\city{Hefei}, \postcode{230088}, \state{Anhui}, \country{China}}}

\abstract{Quantum memory is a core enabling device for large-scale quantum networks, serving as the fundamental building block for quantum repeaters and transportable quantum memories. Among various physical platforms, rare-earth-ion-doped crystals stand out as one of the most promising solid-state candidates for long-lived photonic quantum memories, owing to their exceptionally long spin coherence time and their compatibility with integration. In this review, we first introduce the physical mechanisms underlying long-lived spin coherence in rare-earth systems, with particular emphasis on the zero-first-order-Zeeman transition and dynamical decoupling techniques. We then review the experimental progress in this field, highlighting three distinct lifetimes—the spin coherence time, the light storage time, and the quantum storage time—and covering experiments that have achieved hour-scale spin coherence, minute- to hour-scale light storage, and minute-scale single-photon-level quantum storage. Finally, we discuss the remaining challenges that hinder the realization of practical long-lived quantum memories, explore potential strategies to overcome these obstacles, and outline the prospects for their future applications in quantum networks.
}

\keywords{quantum memories, rare-earth-ions, quantum repeaters, quantum networks}



\maketitle

\section{Introduction}\label{sec1}

Quantum memories (QMs) are devices that enable the coherent storage and on-demand retrieval of photonic quantum states. By converting flying photonic qubits into stationary matter excitations, QMs act as a temporal interface between probabilistic photon transmission and local quantum operations. This capability is essential for synchronizing asynchronous quantum processes and thus represents a key building block for large-scale quantum networks.

\begin{figure*}[t]
\centering
\begin{subfigure}[t]{0.6\textwidth}
    \caption{}
    \centering
    \includegraphics[width=\linewidth]{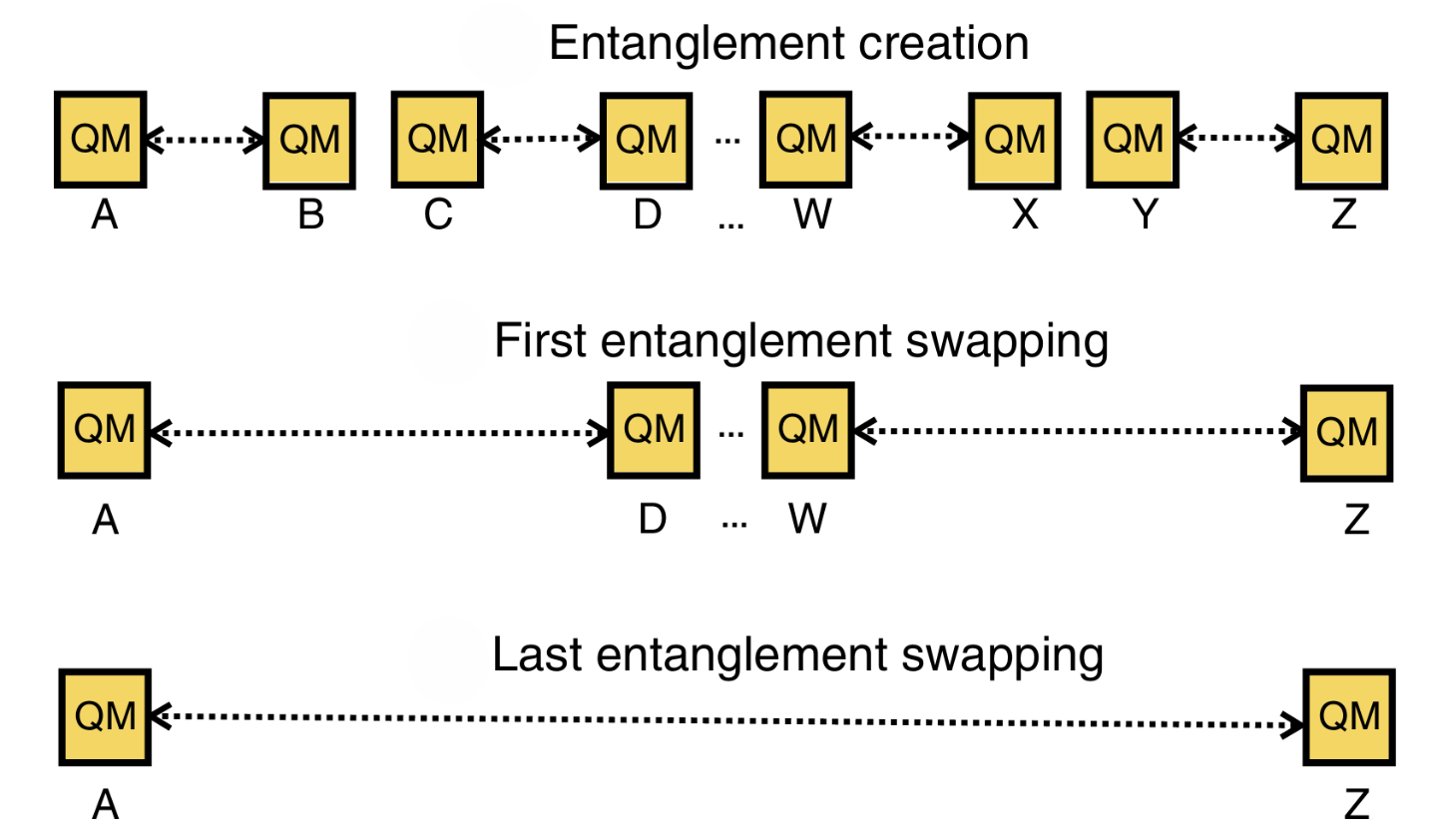}
    
\end{subfigure}
\begin{subfigure}[t]{0.39\textwidth}
    \caption{}
    \centering
    \includegraphics[width=\linewidth]{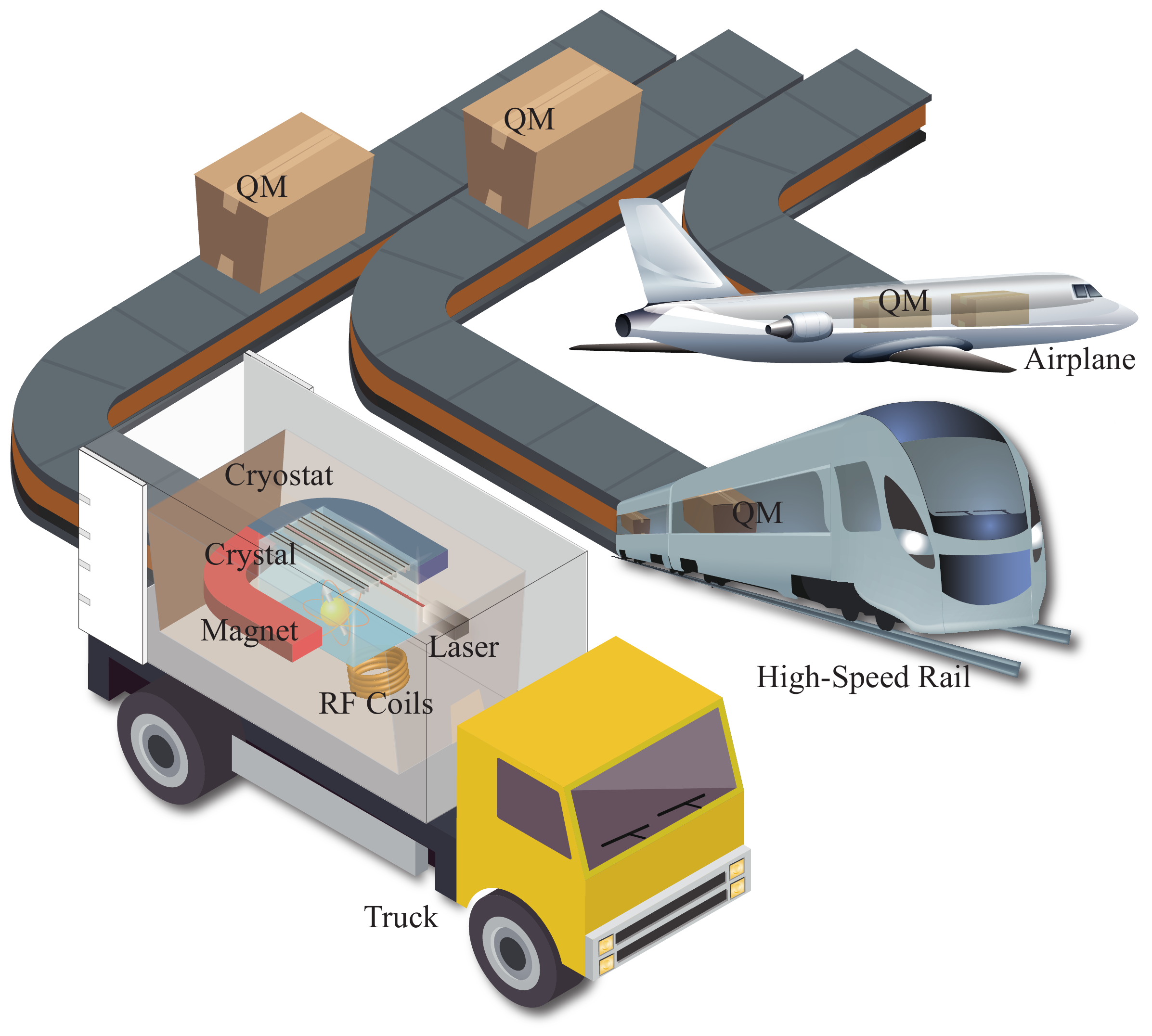}
    
\end{subfigure}

\caption{Applications of long-lived quantum memories in quantum networks. (a) Quantum repeaters dividing the long-distance channel into short segments, generating entanglement between quantum memories in each segment, and then connecting them via entanglement swapping. Reproduced with permission~\cite{RevModPhys.83.33}. Copyright 2011, APS. (b) Transportable quantum memories enabling quantum state transfer between remote nodes via conventional transportation systems. Reproduced with permission~\cite{https://doi.org/10.1002/lpor.202300257}. Copyright 2023 John Wiley and Sons.}\label{fig1}
\end{figure*}

One of the most important applications of QMs is long-distance quantum communication. Owing to exponential photon loss, the direct transmission distance of photons through optical fibers is fundamentally limited to hundreds of kilometers \cite{8267035,RevModPhys.92.025002,PhysRevLett.121.190502,PhysRevLett.124.070501}. Quantum repeaters provide a viable route to overcoming this limitation by dividing a long channel into shorter elementary links, generating entanglement in each link, and connecting the stored entanglement through entanglement swapping \cite{PhysRevLett.81.5932,RevModPhys.83.33,YUAN20101,Muralidharan2016,Zhu2026,Li_2024}, as illustrated in Fig.~\ref{fig1}(a). In such architectures, the memory lifetime must be sufficiently long to accommodate the waiting time required for heralding signals and subsequent swapping operations \cite{RevModPhys.83.33}. Depending on the repeater architecture and the network scale, the required storage time can vary from milliseconds to minutes. An even more stringent requirement arises in the context of transportable quantum memories \cite{Morton2015}, where the quantum memories and their stored quantum information are physically transported between remote nodes using vehicles, drones, or space-based platforms, as sketched in Fig.~\ref{fig1}(b). In this case, the relevant timescale may extend to hours. Moreover, long-lived QMs could also benefit other applications, such as long-baseline optical interferometry \cite{PhysRevLett.109.070503,Bland_Hawthorn_2021,qpzn-h7p9,Stas2026} and quantum computing \cite{PhysRevLett.127.140503,Gottesman1999,Main2025,47zq-2h7t,hvsx-cx2d,Liu2024}.

Many physical systems have been explored for long-lived quantum storage, including atomic ensembles \cite{Eisaman2005,doi:10.1126/science.1140300}, single atoms \cite{vanLeent2022,Specht2011} and rare-earth-ion doped crystals (REICs) \cite{deRiedmatten2008,Hedges2010,Liu2021,Lago-Rivera2021,Zhu2026}. 
REICs stand out as promising solid-state candidates for natural coherence protection, easy integration and scalable applications. 
In REICs, the outer \(5s\) and \(5p\) orbitals shield the \(4f\) electrons, enabling narrow optical transitions and reduced sensitivity to the host environment \cite{Bottger2003}. Selected hyperfine transitions can also exhibit exceptionally long coherence under appropriate magnetic-field conditions \cite{Zhong2015,PRXQuantum.6.010302}. These properties make REICs a natural platform for long-lived optical quantum memories, and in addition, offer excellent compatibility for integration with optical cavities, waveguides, and photonic circuits, which is crucial for scalable applications \cite{https://doi.org/10.1002/lpor.202300257,PhysRevApplied.14.054071,doi:10.1126/science.aan5959,PhysRevLett.123.080502,Saglamyurek2011,Meng2026}.

Here we provide a concise review on the experimental progress in long-lived quantum memories based on REICs, tracing the progression from spin coherence and light storage to quantum storage. Section~\ref{sec2} introduces the main physical approaches for extending storage lifetime, with particular emphasis on the zero-first-order-Zeeman (ZEFOZ) transition and dynamical decoupling (DD). Section~\ref{sec3} reviews representative experimental milestones, including minute-scale classical optical storage, hour-scale nuclear spin coherence time, hour-scale coherent light storage, and minute-scale single-photon-level inputs storage. Section~\ref{sec4} discusses the remaining challenges, such as the trade-off between lifetime and efficiency, noise induced by long DD sequences, and the engineering requirements for integrated and transportable quantum memories.

\section{Approaches to Long-Lived Quantum Memories}
\label{sec2}
Since optical coherence lifetimes are extremely short, long-lived photonic storage requires two key ingredients: the conversion of optical excitations into long-lived spin excitations and the mitigation of decoherence on the spin transition.

The optical-to-spin conversion can be implemented using several quantum memory protocols, including electromagnetically induced transparency (EIT) \cite{PhysRevLett.84.5094}, controlled reversible inhomogeneous broadening (CRIB) \cite{PhysRevLett.87.173601}, atomic frequency comb (AFC) \cite{PhysRevA.79.052329}, and noiseless photon echo (NLPE) \cite{Ma2021}. Among these, AFC and NLPE have successfully enabled spin-wave quantum storage in REICs, as will be discussed in detail in Section 3.

After optical-to-spin conversion, environmental magnetic fluctuations cause decoherence of the spin transition. The ZEFOZ technique and DD mitigate this decoherence in complementary ways: ZEFOZ reduces the intrinsic magnetic-field sensitivity of a selected transition of the target spins, whereas DD averages out time-dependent environmental noise through controlled refocusing of the target spins. Here we describe these two techniques in more detail.

\subsection{ZEFOZ}
In REICs, random flips of the host nuclear spins and fluctuations of the laboratory magnetic field can perturb the transition frequency of target rare-earth ions via the Zeeman effect. Under a specific magnetic-field magnitude and orientation, the first-order derivative of the selected ZEFOZ transition frequency vanishes, providing straightforward coherence protection.

The transition frequency between two nuclear-spin sublevels, $\ket{k}$ and $\ket{l}$, denoted as $\omega_{kl}(\bm{B})$, can be expanded in a Taylor series around an external DC magnetic field $\bm{B}_0$ as follows:
\begin{equation}
\omega_{kl}(\bm{B}) = \omega_{kl}(\bm{B}_0) + \bm{S}_1 \cdot \Delta \bm{B} + \frac{1}{2} \Delta \bm{B} \cdot \bm{S}_2 \cdot \Delta \bm{B},
\end{equation}
where $\Delta \bm{B} = \bm{B} - \bm{B}_0$ denotes the ambient magnetic-field fluctuations originating from the nuclear spins in the host lattice; $\bm{S}_1$ is the first-order Zeeman gradient vector, describing the linear response of the transition frequency to magnetic perturbations; and $\bm{S}_2$ is the second-order Zeeman curvature tensor, describing the quadratic response.

For a ZEFOZ transition, the spin coherence time $T_2$ can be roughly estimated as
\begin{equation}
\frac{1}{T_2}\sim \left\lvert\bm{S}_1 \cdot \Delta \bm{B}\right\lvert + \frac{1}{2} \Delta \bm{B} \cdot \bm{S}_2 \cdot \Delta \bm{B}.
\end{equation}
The cancellation of $\bm {S}_1$ can extend the spin coherence time by orders of magnitude. 

In practice, however, ensemble inhomogeneity, the second-order curvature, and imperfect DC field alignment can still introduce measurable decoherence. Consequently, identifying high-quality ZEFOZ transitions remains the most crucial task for realizing long-lived quantum memories.

\subsection{Dynamical decoupling (DD)}
Limited by the above-mentioned imperfections, ZEFOZ transitions can only achieve a Hahn-echo spin coherence time $T_2$ of minutes. To further extend it, one can employ DD by rapidly applying control pulses that periodically reverse the phase evolution. The basic idea originates from the Hahn echo \cite{PhysRev.80.580}: a $\pi/2$ pulse creates transverse spin coherence, the spins dephase under local frequency offsets, and a later $\pi$ pulse reverses the phase evolution, so that an echo forms after an equal delay.

DD extends this refocusing principle from a single echo to a pulse train. By applying repeated $\pi$ pulses, the system filters out noise over a wider bandwidth and becomes less sensitive to environmental disturbances. Key variants include the CPMG sequence \cite{10.1063/1.1716296}, which provides self-compensation of pulse flip-angle errors by rotating the phase of the $\pi$ pulse by $90^\circ$; the Uhrig sequence \cite{PhysRevLett.98.100504}, which optimizes non-uniform pulse timing for superior suppression of $1/f$-type low-frequency noise; and sequences such as XY4 \cite{GULLION1990479} and KDD \cite{PhysRevLett.106.240501}, which are designed to suppress errors arising from imperfect pulses, including flip-angle errors and detuning.

Extending the spin coherence time $T_2$ to the hour scale may require thousands of $\pi$ pulses. Even small errors in pulse area, phase, or detuning can accumulate into significant population noise. This poses a major challenge for ensemble-based photonic quantum memories, where the signal-to-noise ratio for single-photon-level inputs is a critical performance metric. In addition, dense microwave or RF pulses can heat the crystal, introducing technical limits on the overall quantum storage performance.

\begin{figure*}[t]
\begin{subfigure}[t]{0.44\textwidth}
    \caption{}
    \centering
    \includegraphics[width=\linewidth]{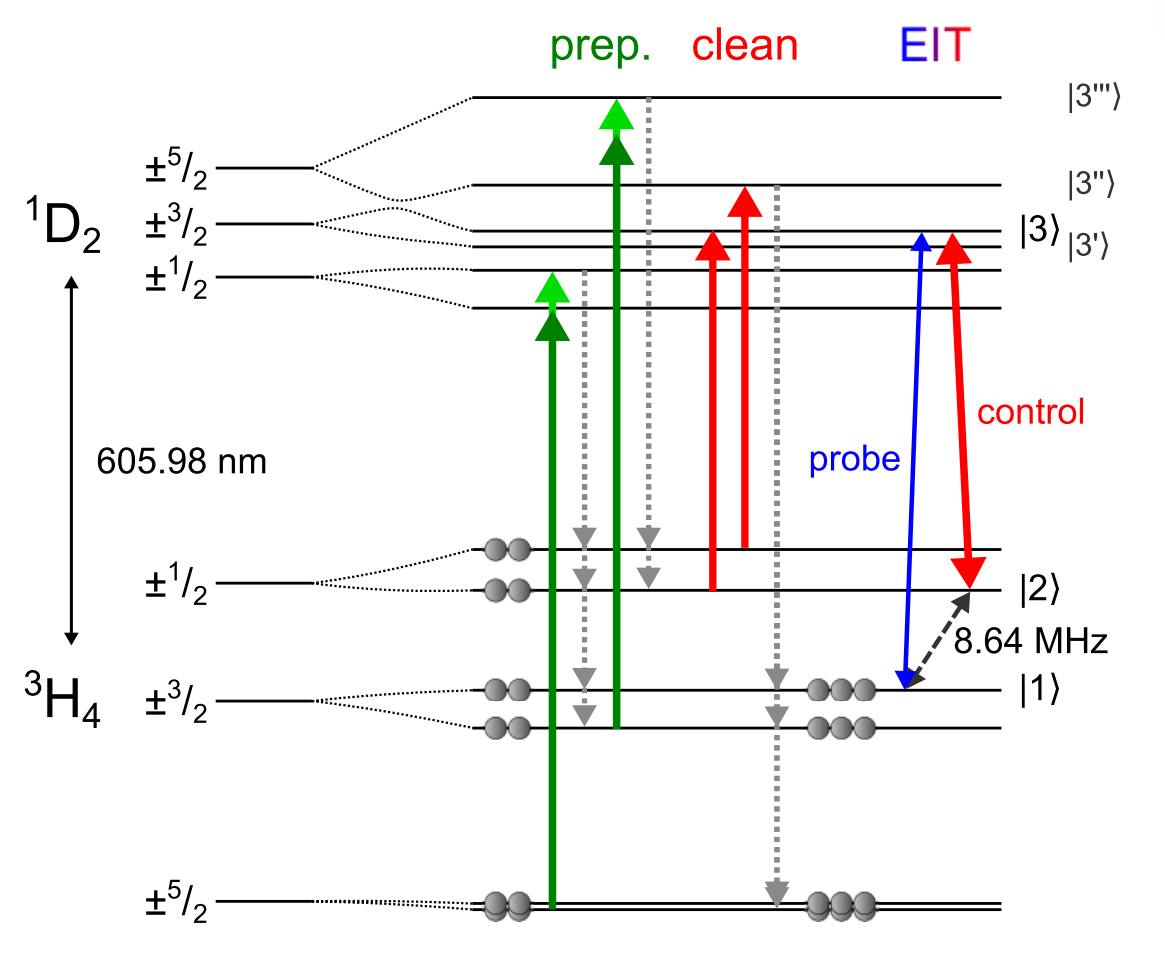}
    
\end{subfigure}
\begin{subfigure}[t]{0.55\textwidth}
    \caption{}
    \centering
    \includegraphics[width=\linewidth]{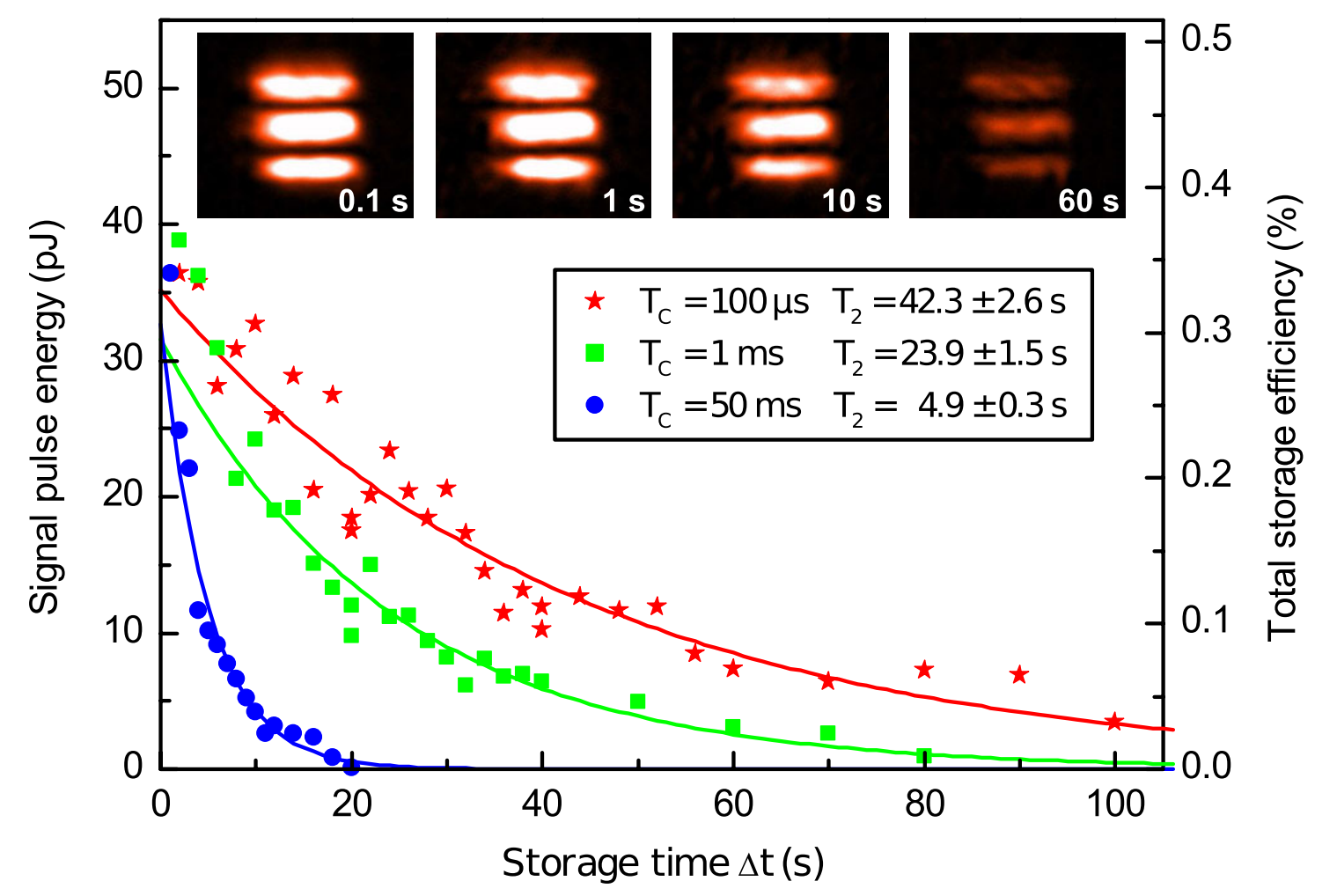}
\end{subfigure}
\caption{Light storage for 1 minute in Pr$^{3+}$:Y$_2$SiO$_5$ crystals. (a) Energy-level scheme for EIT storage using a ZEFOZ transition in Pr$^{3+}$:Y$_2$SiO$_5$ crystals. Optical pumping (green) and background‑cleaning (red) pulses prepare all ions in state $\ket{1}$. A strong control beam (red) then couples the $\ket{2}\leftrightarrow\ket{3}$ transition, while a weak probe beam (blue) couples the $\ket{1}\leftrightarrow\ket{3}$ transition. Reversible mapping of photonic quantum states onto atomic spin-wave coherence is achieved by switching the control beam on and off. (b) Decay of signal pulse energy over storage time under DD sequences with variable pulse intervals. Reproduced with permission~\cite{PhysRevLett.111.033601}. Copyright 2013, APS.}
\label{fig2}
\end{figure*}

\section{Experimental Progress of Long-lived Photonic Quantum Memories}
\label{sec3}

Building on the coherence protection techniques introduced in the previous section, the storage lifetime of REICs has been improved by orders of magnitude over the past decade. The progress of REIC-based long-lived quantum memories can be understood through three related but distinct lifetimes.

\textbf{The spin coherence time} measures how long an ensemble spin remains phase coherent and sets an upper bound for the light storage time. \textbf{The light storage time} is the duration for which an optical signal can be mapped to a spin excitation and later retrieved with phase coherence. In practice, it is bounded by the spin coherence time and further limited by experimental conditions. For example, balancing efficiency may require increasing the ion doping concentration in the crystal, which in turn intensifies inter-ion coupling, or lengthening the crystal, which could reduce the achievable spin coherence time. \textbf{The quantum storage time} is the duration for which a single-quantum-level signal can be stored, bounded by the light storage time. For such weak inputs, the storage process must achieve a signal-to-noise ratio (SNR) far above unity, and the retrieved state must retain a fidelity above that of the best classical strategy. This SNR requirement limits the achievable storage time, since long lifetimes often require a large amount of DD pulses, which are accompanied by significant noise \cite{Ortu2022,doi:10.1126/sciadv.adu5264,lv2025minutescalephotonicquantummemory}.

The field of long-lived photonic memories has developed in stages. Early work established minute-scale storage of classical optical fields. Later experiments demonstrated hour-scale spin coherence and then achieved hour-scale coherent optical storage. The current frontier is single-photon-level quantum storage in the minute regime, where noise becomes the decisive constraint.

\subsection{Minute-Scale Coherent Light Storage using EIT}

The ZEFOZ transition was first identified in Pr$^{3+}$:Y$_2$SiO$_5$ crystals. At a critical magnetic field of \SI{78}{\milli\tesla}, an ensemble spin $T_2$ of 27.9 s was achieved using DD with a pulse interval of 7.5 ms \cite{PhysRevLett.95.030506}. Subsequently, storage of classical light was demonstrated using the EIT protocol (see Fig.~\ref{fig2}(a)), reporting a $1/e$ light storage time of 2.3 s with a DD interval of 100 ms \cite{PhysRevLett.95.063601}. Reducing the interval to \(\SI{100}{\micro\second}\) extended the fitted $1/e$ light storage time to \(\SI{42.3}{\second}\), with coherent light and images retrieved after 1 minute (see Fig.~\ref{fig2}(b)) \cite{PhysRevLett.111.033601}. This progression illustrates the lifetime gain from faster refocusing, while also anticipating the pulse-error and heating constraints of longer sequences. In 2025, EIT storage of single-photon-level inputs for 1 s was achieved in Pr$^{3+}$:Y$_2$SiO$_5$ crystals \cite{Hain_2025}, but the reported SNR is insufficient for photonic qubit storage. At a short storage time of \SI{2}{\micro\second} (without RF rephasing pulses), the single-photon SNR was \(2.2 \pm 0.2\); at \SI{1}{\second}, it decreased to \(1.3 \pm 0.3\). The main difficulty of EIT quantum storage is that the strong control field creates a large optical background overlapping the retrieved signal.

The nuclear spin population lifetime $T_1$ of Pr$^{3+}$:Y$_2$SiO$_5$ crystals is 145 s \cite{PhysRevLett.95.030506}, which imposes a fundamental limit on the achievable light storage time. The minute-scale light storage already approaches the fundamental limit of this material system.

\begin{figure*}[t]
\begin{subfigure}[t]{0.49\textwidth}
    \caption{}
    \centering
    \includegraphics[width=\linewidth]{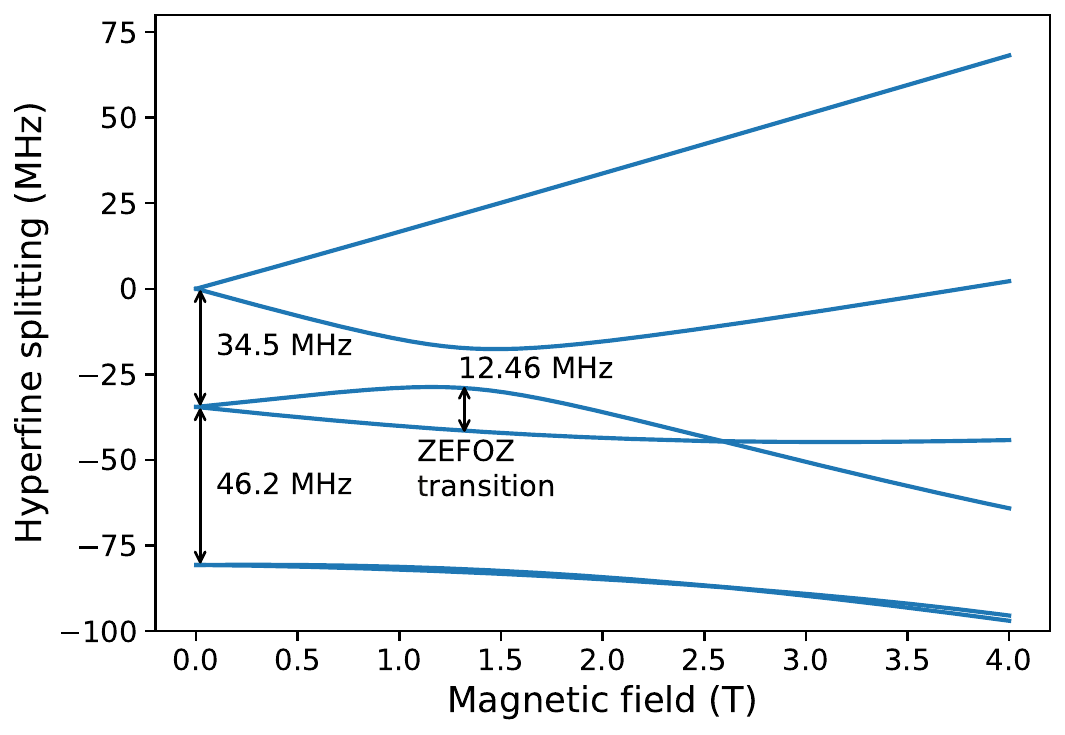}
\end{subfigure}
\begin{subfigure}[t]{0.49\textwidth}
    \caption{}
    \centering
    \includegraphics[width=\linewidth]{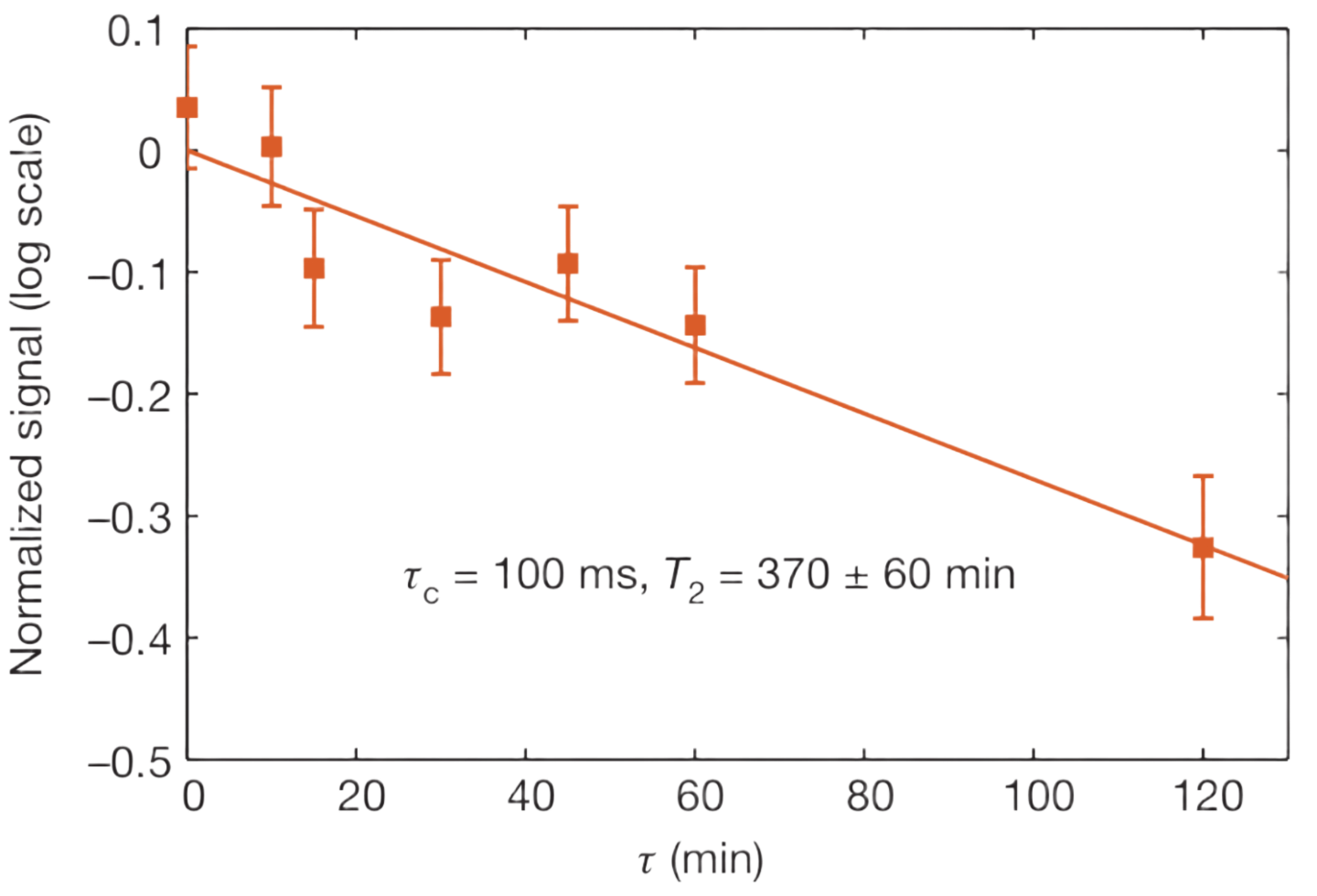}
\end{subfigure}

\caption{Spin coherence time of 6 hours in Eu$^{3+}$:Y$_2$SiO$_5$ crystals. (a) A ZEFOZ transition commonly used in Eu$^{3+}$:Y$_2$SiO$_5$ crystals. With a magnetic field of \SI{1.3}{\tesla} applied along the direction $[-0.535,-0.634,0.558]$ in the crystal's $D_1 \times D_2 \times b$ coordinate frame, the first-order Zeeman coefficient vanishes for the indicated ZEFOZ transition, yielding a Hahn-echo spin $T_2$ of 47 s. We note that the level structure presented here is an accurate level diagram calculated with the effective Hamiltonian from Ref.~\cite{MA201832}, whereas the original level diagram in Ref.~\cite{Zhong2015} contained an incorrect ordering. (b) By applying a KDD$_x$ sequence with a pulse interval of 100 ms, the spin $T_2$ under DD can be further extended to approximately 6 hours. Reproduced with permission~\cite{Zhong2015}. Copyright 2015, Springer Nature.}
\label{fig3}
\end{figure*}

\subsection{Hour-Scale Nuclear Spin Coherence Time}

To overcome the $T_1$ limit of Pr$^{3+}$:Y$_2$SiO$_5$ crystals, Eu$^{3+}$:Y$_2$SiO$_5$ crystals have been extensively studied, since their spin $T_1$ can approach 23 days at 2 K \cite{PhysRevB.68.085109}. According to the hyperfine-interaction Hamiltonian for Eu$^{3+}$ ions \cite{MA201832,PhysRevB.74.195101,PhysRevB.97.094416}, the second-order Zeeman curvature of the Eu$^{3+}$ ground-state hyperfine levels (\SI{21.7}{\mega\hertz\per\tesla\squared}) is two orders of magnitude lower than that of Pr$^{3+}$, suggesting that a much longer $T_2$ can be expected.

In 2015, Zhong \textit{et al.} experimentally determined the ZEFOZ transition for $^{151}$Eu$^{3+}$ ions doped in Y$_2$SiO$_5$ crystals at a critical magnetic field of 1.3 T (see Fig.~\ref{fig3}(a))~\cite{Zhong2015}. The ZEFOZ transition achieves a Hahn-echo $T_2$ of 47 s, with contributions from both the ZEFOZ effect and the ``frozen-core'' effect: the strong magnetic field induces a magnetic moment on the Eu$^{3+}$ ions, which produces a $\sim$kHz dipolar interaction with neighboring Y$^{3+}$ nuclear spins \cite{Zhong_2019}. This interaction reduces the local Y spin-flip rate by orders of magnitude compared to that with zero field, significantly reducing the magnetic noise of Y spin bath.

Finally, an ensemble spin $T_2$ of $370 \pm \SI{60}{\minute}$ ($\sim$6 hours) is obtained by further applying a DD sequence based on KDD$_x$ with a pulse interval of \SI{100}{\milli\second} (see Fig.~\ref{fig3}(b)). This experiment reported the longest known spin $T_2$ among all materials to date and indicated that Eu$^{3+}$:Y$_2$SiO$_5$ crystals are a promising candidate for optical quantum memories with ultralong lifetimes.

\subsection{Hour-Scale Storage of Light using AFC}
Although a long-lived spin $T_2$ has been obtained in Eu$^{3+}$:Y$_2$SiO$_5$ crystals, this spin-coherence result did not immediately yield optical storage because the complete energy level structure at the ZEFOZ field remained unknown.
In 2018, two independent works successfully determined the hyperfine Hamiltonian of the optically excited state for Eu$^{3+}$ ions \cite{MA201832,PhysRevB.97.094416}, enabling a rough prediction of the level structure at the ZEFOZ field. 

\begin{figure}[!h]

\centering
\includegraphics[width=0.9\linewidth]{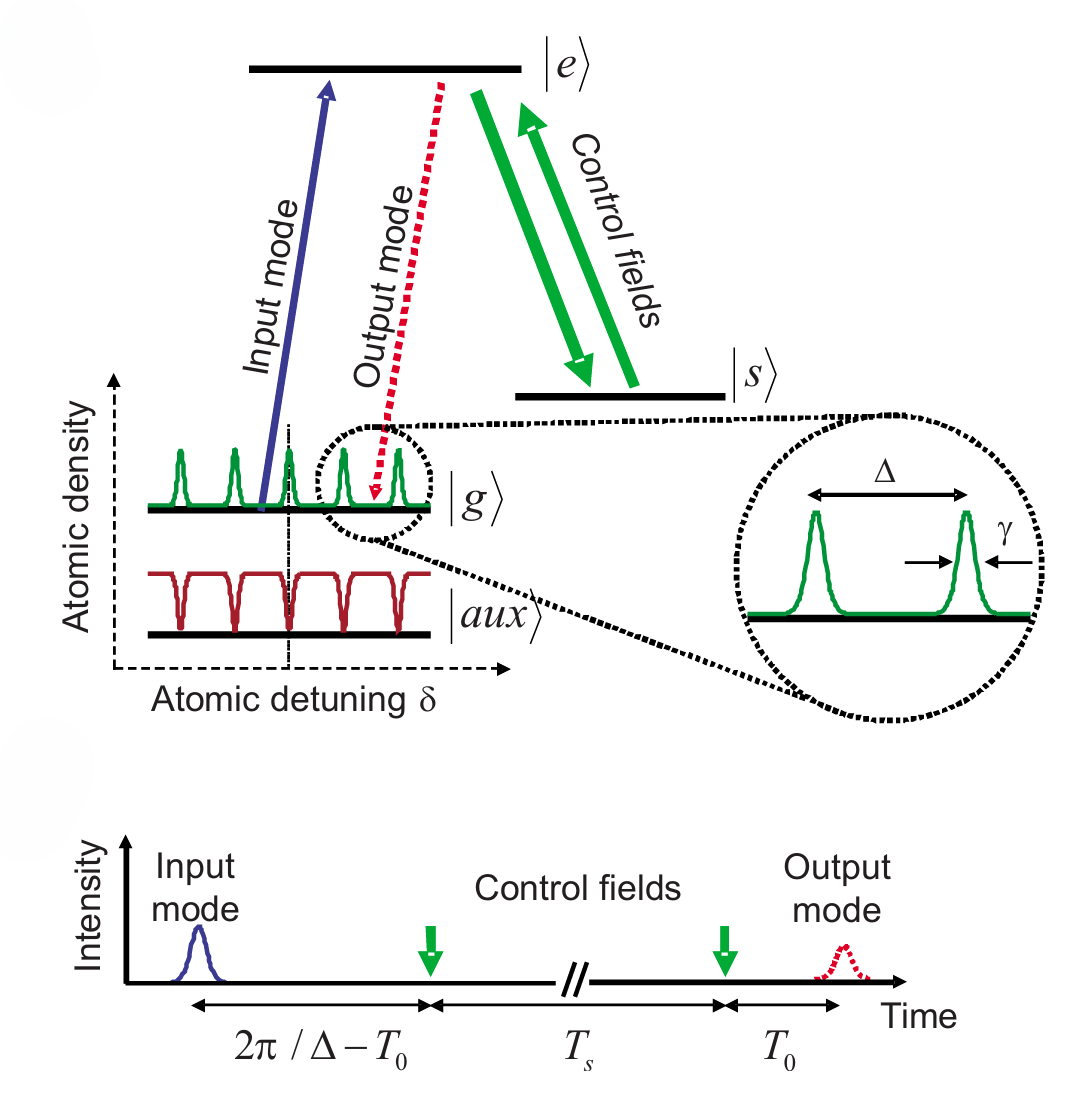}

\caption{Atomic frequency comb (AFC) quantum memory protocol. The AFC is prepared on the $\ket{g}$ level via spectral hole burning, while the $\ket{s}$ level is left empty. The input photon is absorbed by the AFC and read out at time $T_s + 2\pi/\Delta$ in the presence of the control fields. Reproduced with permission~\cite{PhysRevA.79.052329}. Copyright 2009, APS.}
\label{fig4}
\end{figure}

\begin{figure*}[!h]
    \centering

    \begin{subfigure}[t]{0.46\textwidth}
    \caption{}
    \centering
    \includegraphics[width=\linewidth]{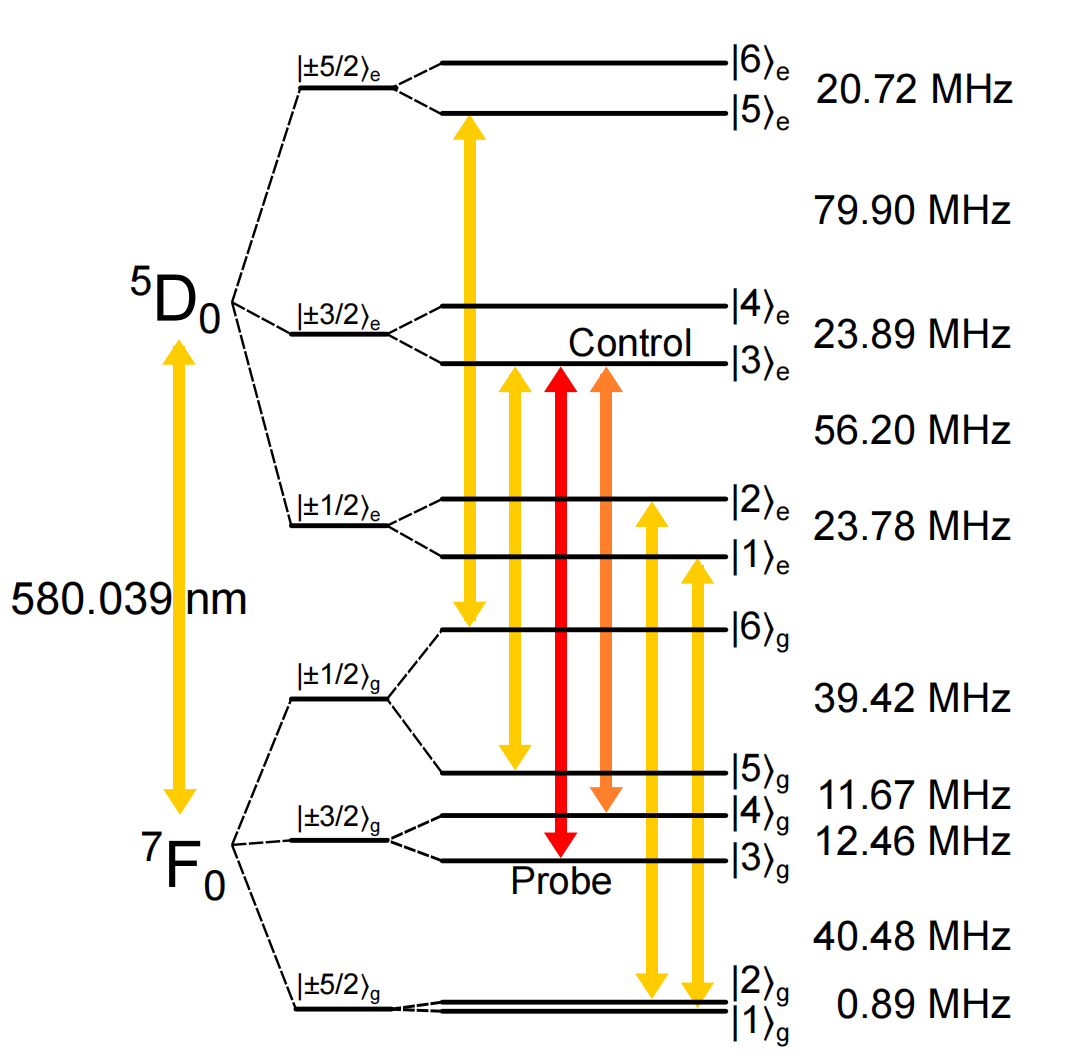}
\end{subfigure}\begin{subfigure}[t]{0.53\textwidth}
    \caption{}
    \centering
    \includegraphics[width=\linewidth]{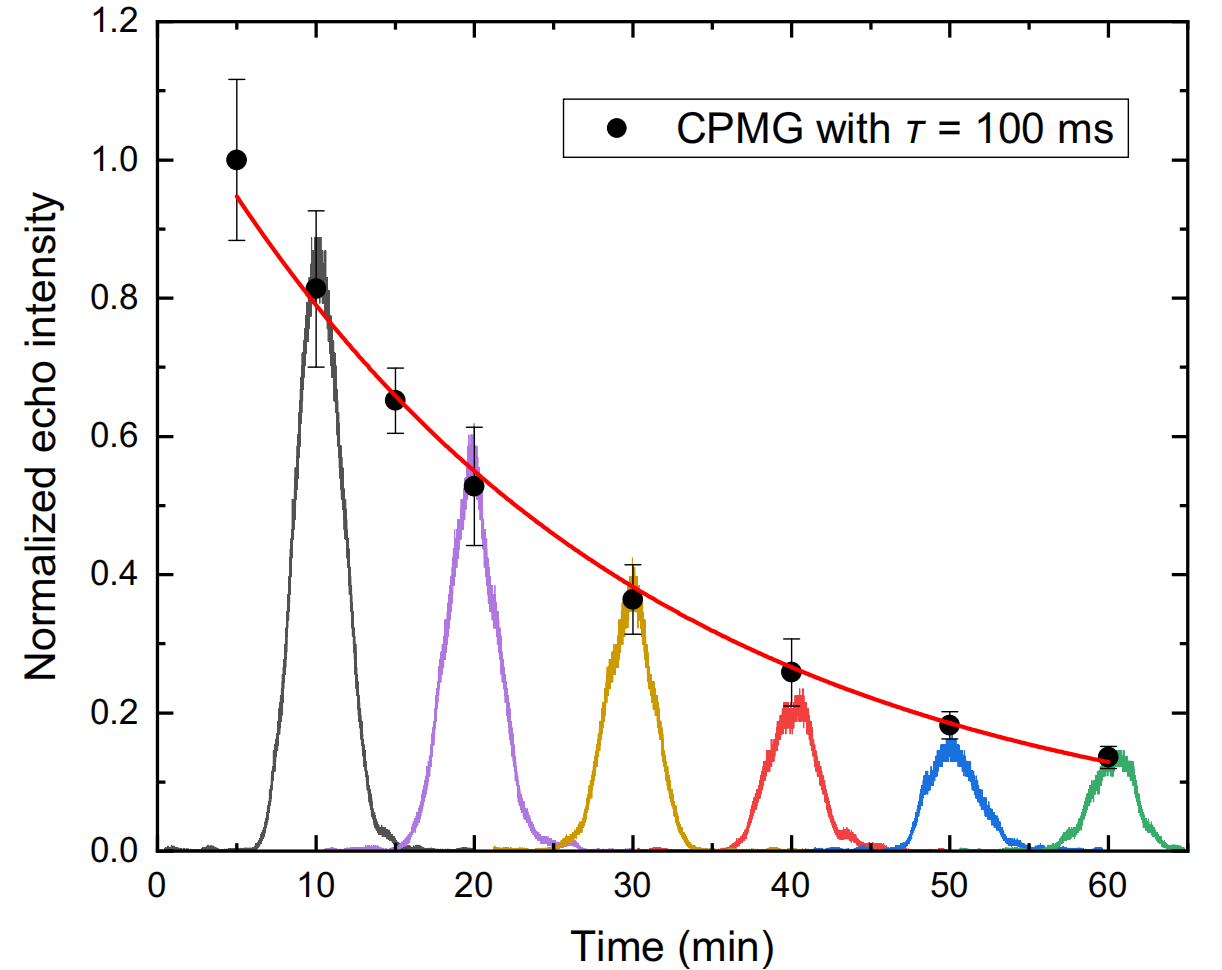}
\end{subfigure}

\caption{Coherent light storage for 1 hour in $^{151}$Eu$^{3+}$:Y$_2$SiO$_5$ crystals. (a) Level structure of $^{151}$Eu$^{3+}$ at the \SI{1.3}{\tesla} ZEFOZ field. The AFC is prepared on state $\ket{3}_g$, and the states $\ket{3}_g$, $\ket{4}_g$, and $\ket{3}_e$ constitute a $\Lambda$ system for light storage. (b) Storage of light for 1 hour using the CPMG sequence for DD with a pulse interval of \SI{100}{\milli\second}. The $1/e$ lifetime is measured to be $52.9 \pm \SI{1.2}{\minute}$. Reproduced under the terms of the CC BY 4.0 license~\cite{Ma2021(1)}. Copyright 2021, Springer Nature.}
    \label{fig:8}
\end{figure*}

In 2021, Ma \textit{et al.} experimentally determined the complete level structure, as shown in Fig.~\ref{fig:8}(a), using pulsed Raman-heterodyne-detected NMR spectroscopy~\cite{Ma2021(1)}. 
A $\Lambda$-type three-level system consisting of $\ket{3}_g$, $\ket{4}_g$, and $\ket{3}_e$ was identified, making it possible to implement optical quantum memory protocols for spin-wave storage of light. 
This experiment is based on the AFC protocol \cite{PhysRevA.79.052329}, with the basic pulse sequence shown in Fig.~\ref{fig4}. 
The AFC is prepared via spectral hole burning, with a bandwidth of $\Gamma$ and a period of $\Delta$. Absorption of an input mode resonant with the $\ket{g}\leftrightarrow\ket{e}$ transition creates the collective state:
\begin{equation}
    \ket{\psi}=\sum_{j=1}^N c_j e^{i\delta_j t} e^{-ikz_j}\ket{g_1\cdots e_j\cdots g_N}.
\end{equation}
Owing to the atomic phase evolution $e^{i\delta_j t}$, the ensemble quickly dephases but rephases at $t = 2\pi/\Delta$ due to the periodic structure of the AFC. 
Using a control pulse with a pulse area of $\pi$, resonant with the excited state and the spin state $\ket{s}$ (corresponding to $\ket{4}_g$ in Fig.~\ref{fig:8}(a)), the optical excitation can be transferred to a spin-state excitation. After applying DD to extend the spin $T_2$ for long-term storage, an additional control pulse converts the spin-wave excitation back to an optical echo emission. 

Using a DD sequence based on CPMG with a pulse interval of \SI{100}{\milli\second}, this experiment achieved a $1/e$ light storage time of $52.9 \pm \SI{1.2}{\minute}$ (see Fig.~\ref{fig:8}(b)), representing a 75-fold improvement over the previous record for the light storage time \cite{PhysRevLett.111.033601}. A time-bin-like interference measurement verified coherent light storage after 1 hour, with a visibility of $92.9 \pm 4.9\%$.

Through this successful combination of ZEFOZ, AFC, and DD techniques, this work successfully translated hour-scale spin coherence into hour-scale coherent optical storage (Fig.~\ref{fig:8}(b)). To date, Eu$^{3+}$:Y$_2$SiO$_5$ remains the only material system that has demonstrated coherent light storage in the hour regime, underscoring the feasibility of implementing transportable quantum memories.

\subsection{Minute-Scale storage of single-photon-level inputs using NLPE}

Storage at the single-photon level places stricter demands on efficiency and noise than coherent-light storage. For Fock-state inputs, the fidelity must exceed that of the best classical strategy, which requires an SNR above unity \cite{PhysRevLett.74.1259}. For weak coherent inputs at the single photon level, the SNR needs to be even higher since the classical strategy can make use of the multi-photon events. This demands that the storage process achieves both high efficiency and low noise simultaneously.

\begin{figure*}[!h]
    \centering

    \begin{subfigure}[t]{0.51\textwidth}
    \caption{}
    \centering
    \includegraphics[width=\linewidth]{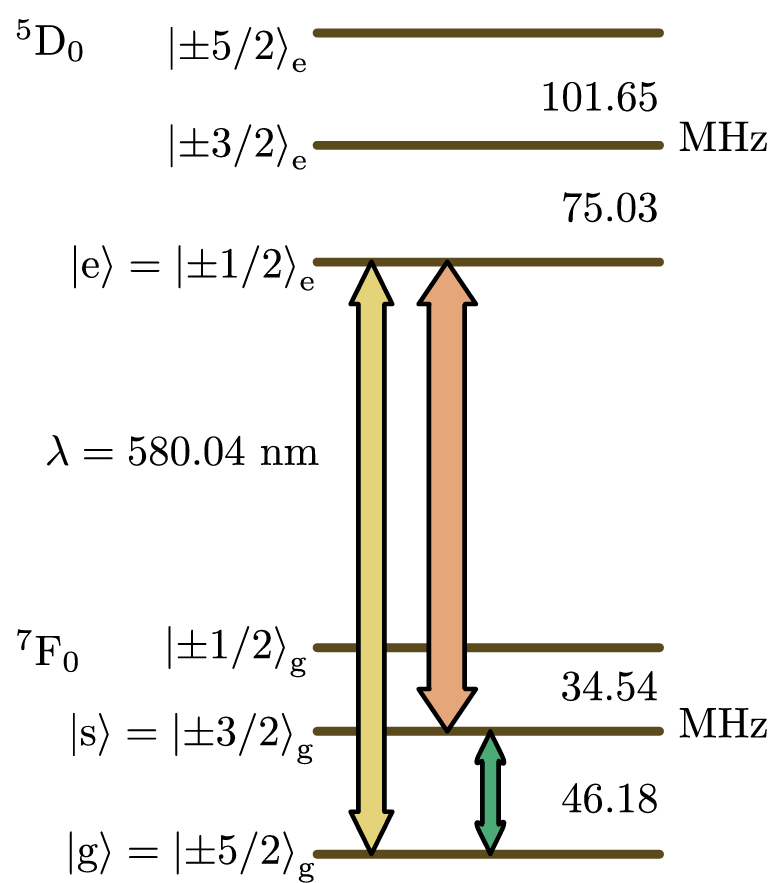}
    
\end{subfigure}\begin{subfigure}[t]{0.48\textwidth}
    
    \centering
    \caption{}
    \includegraphics[width=\linewidth]{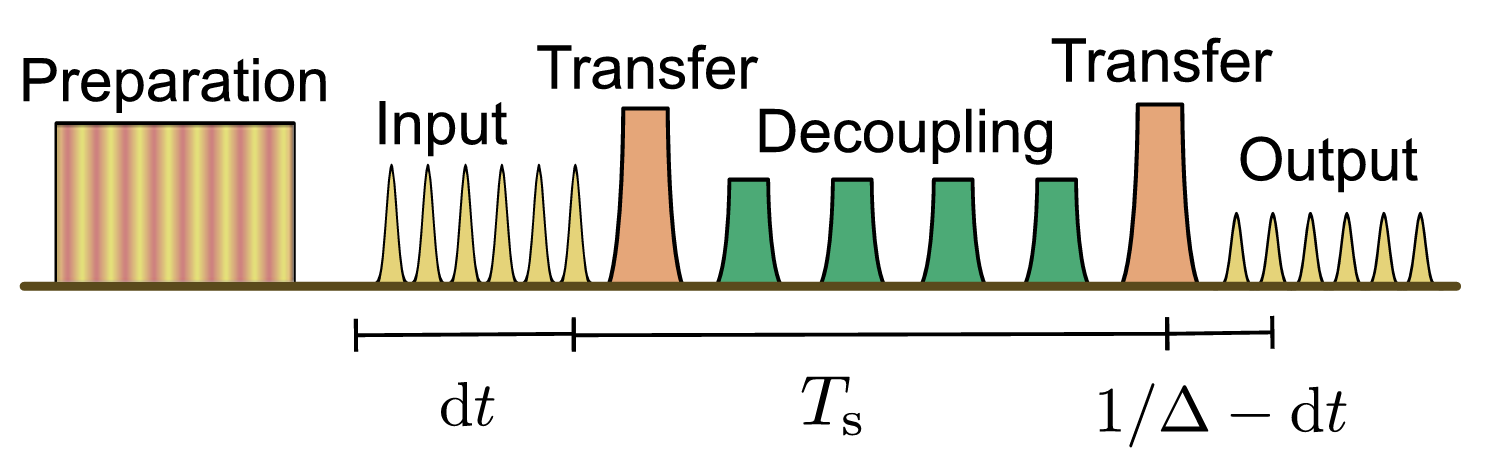}
    \centering
    \caption{}
    \includegraphics[width=\linewidth]{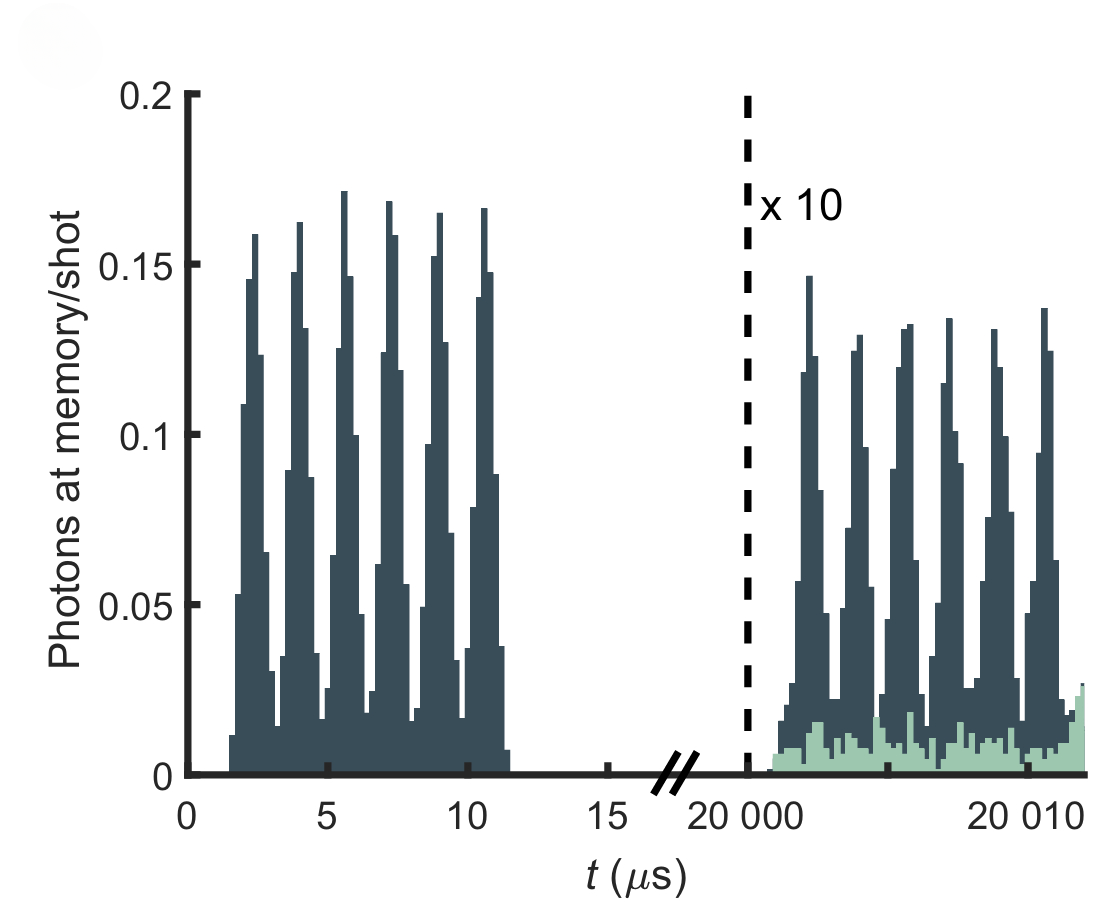}
\end{subfigure}

\caption{Sub-second storage of single-photon-level inputs in a $^{151}$Eu$^{3+}$:Y$_2$SiO$_5$ crystal. (a) Energy-level scheme of $^{151}$Eu$^{3+}$:Y$_2$SiO$_5$, indicating the transitions exploited for storage. (b) Pulse sequence for the spin-wave AFC protocol with DD. (c) Multimode storage of single-photon-level inputs at a storage time of 20 ms, achieving a storage efficiency of $7.39\pm0.04\%$ and an SNR of $7.4\pm0.5$. Reproduced under the terms of the CC BY 4.0 license~\cite{Ortu2022}.}
    \label{fig5}
\end{figure*}

In 2015, AFC and DD were combined at zero magnetic field \cite{PhysRevLett.114.230502}, demonstrating the storage of single-photon-level inputs for 1 ms. In 2022, the storage time was extended to the sub-second regime by applying a weak magnetic field (1.35 mT) to improve the spin coherence time (see Fig.~\ref{fig5}(a) for the energy-level scheme and Fig.~\ref{fig5}(b) for the pulse sequence) \cite{Ortu2022}. As shown in Fig.~\ref{fig5}(c), this achieved an SNR of 7.4 at 20 ms and 2.5 at 100 ms for single-photon-level inputs.

Although AFC has successfully enabled long-lived classical-light storage using ZEFOZ transitions, it suffers from weak absorption and from dilution of the optical depth during comb preparation. Specifically, the initial absorption depth $d_0$ is reduced to $d_0/F_{\mathrm{AFC}}$, where $F_{\mathrm{AFC}}$ is the finesse of the AFC. 
While high efficiency requires strong absorption and thus a high doping concentration, a long spin coherence time requires low ensemble broadening and thus a low doping concentration \cite{PhysRevB.107.014310}. Long-lived quantum memories must therefore balance these competing requirements, representing a trade-off between efficiency and storage time.

\begin{figure}[!h]

\centering
\includegraphics[width=0.9\linewidth]{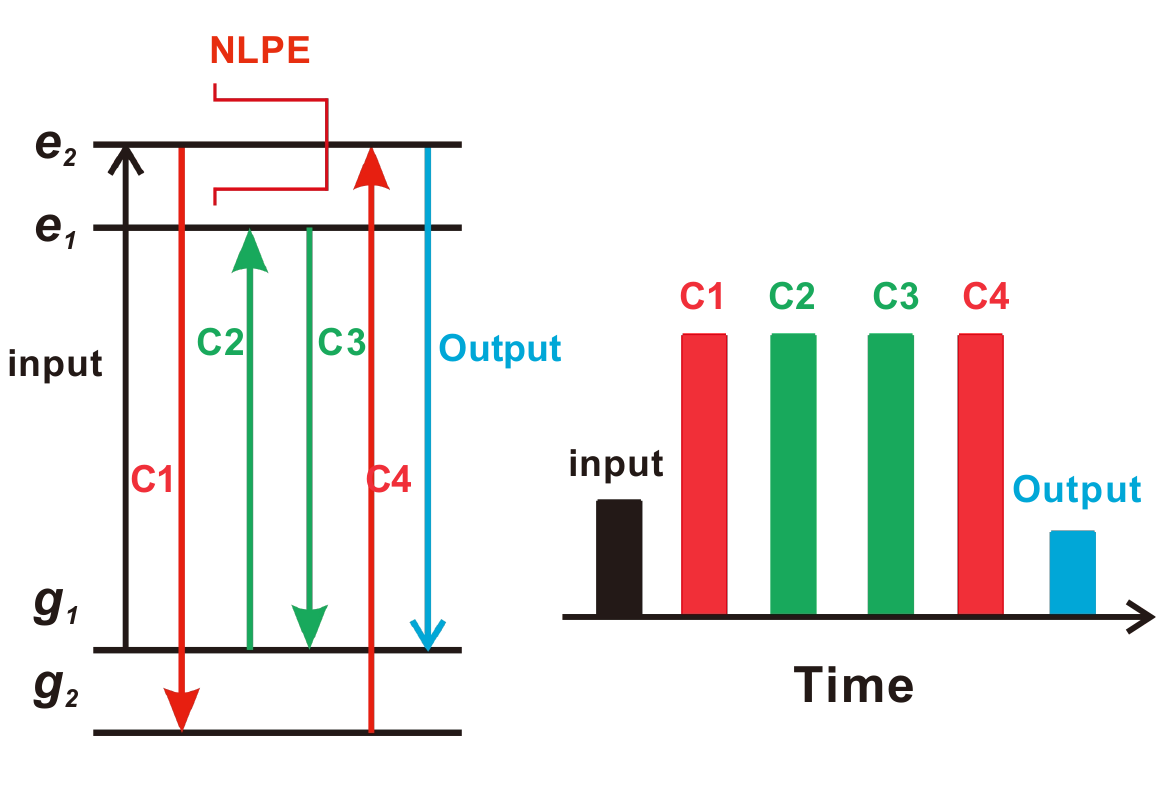}

\caption{NLPE protocol. Input photons are absorbed via the natural absorption prepared on $g_1$. After four optical $\pi$ pulses (C1--C4), the noiseless photon echo is emitted from $e_2$, while the residual population remains in $e_1$. This allows the spontaneous emission noise and the NLPE echo to be frequency-separated, yielding a noise-free storage process. DD can be inserted between C1 and C2, during which the photon is stored as a spin-wave excitation in the ground state. Reproduced with permission~\cite{https://doi.org/10.1002/lpor.202300257}. Copyright 2023 John Wiley and Sons.}
\label{nlpe}
\end{figure}

To achieve efficient optical absorption and storage at ZEFOZ fields, one needs a quantum memory protocol that can realize quantum storage using the natural optical absorption of the material. In 2021, Ma \textit{et al.} proposed the NLPE protocol \cite{Ma2021}, which successfully achieves this goal. Following the principle of standard photon echo, they implemented atomic rephasing using optical $\pi$ pulses so that the natural absorption profile can be directly utilized for photon absorption. The NLPE protocol involves four atomic levels rather than the typical $\Lambda$ system, which is crucial for eliminating the spontaneous emission noise associated with the optical rephasing process. As shown in Fig.~\ref{nlpe}, the coherent NLPE echo emission arises from the $e_2$ state, while the residual population is left in the $e_1$ state. As a result, the signal echo is free from spontaneous emission noise, which had been regarded as a fundamental obstacle to implementing quantum storage using photon-echo techniques \cite{PhysRevA.81.062333,PhysRevA.79.053851}.

In 2025, NLPE and DD were combined at zero magnetic field, achieving photonic qubit storage for 1 ms in a laser-written waveguide fabricated in a $^{151}$Eu$^{3+}$:Y$_2$SiO$_5$ crystal~\cite{doi:10.1126/sciadv.adu5264}. This is the first integrated quantum memory that outperforms a simple fiber delay line in efficiency.

\begin{figure*}[!h]
    \centering

    \begin{subfigure}[t]{0.46\textwidth}
    \caption{}
    \centering
    \includegraphics[width=\linewidth]{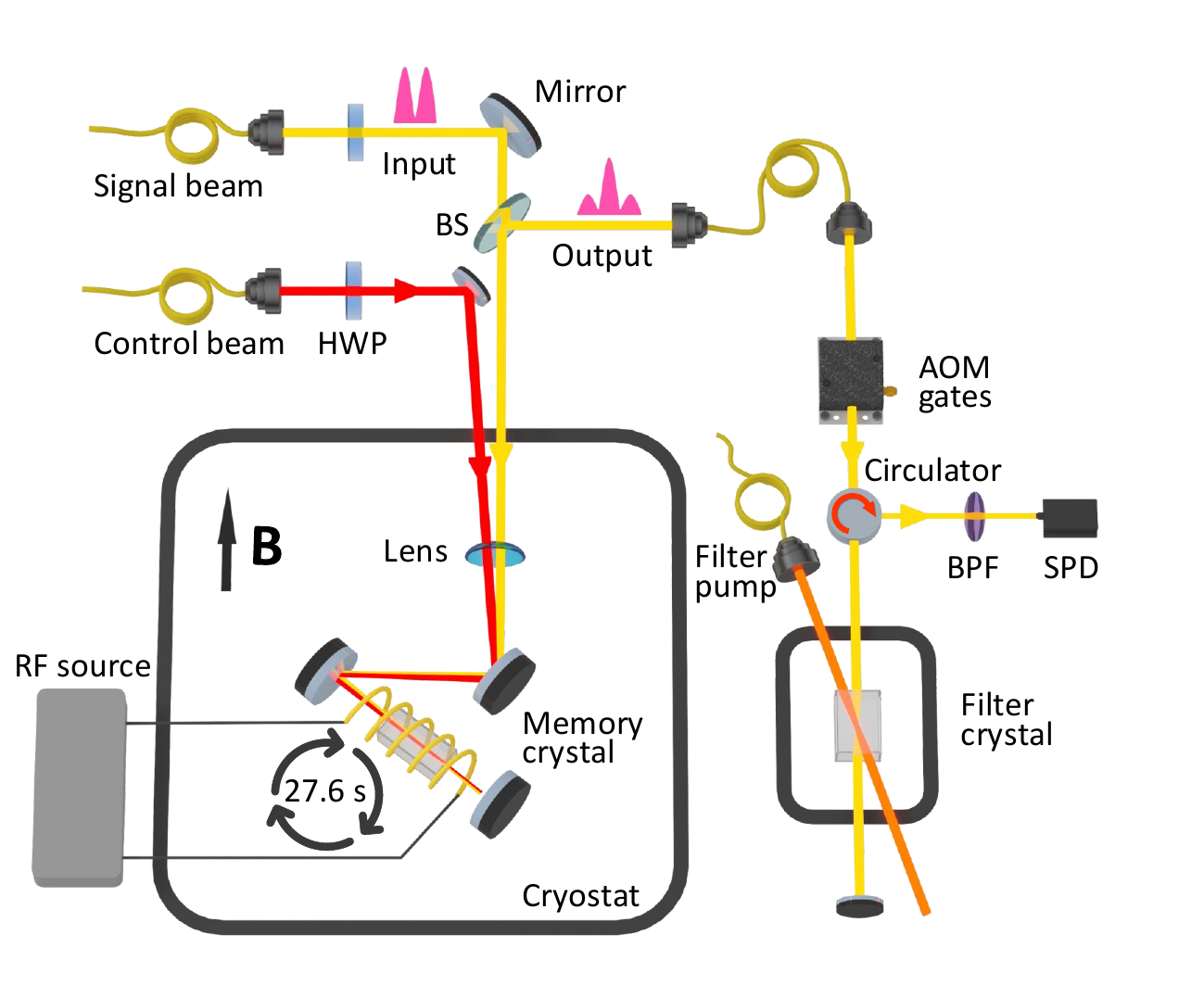}
\end{subfigure}\begin{subfigure}[t]{0.53\textwidth}
    \caption{}
    \centering
    \includegraphics[width=\linewidth]{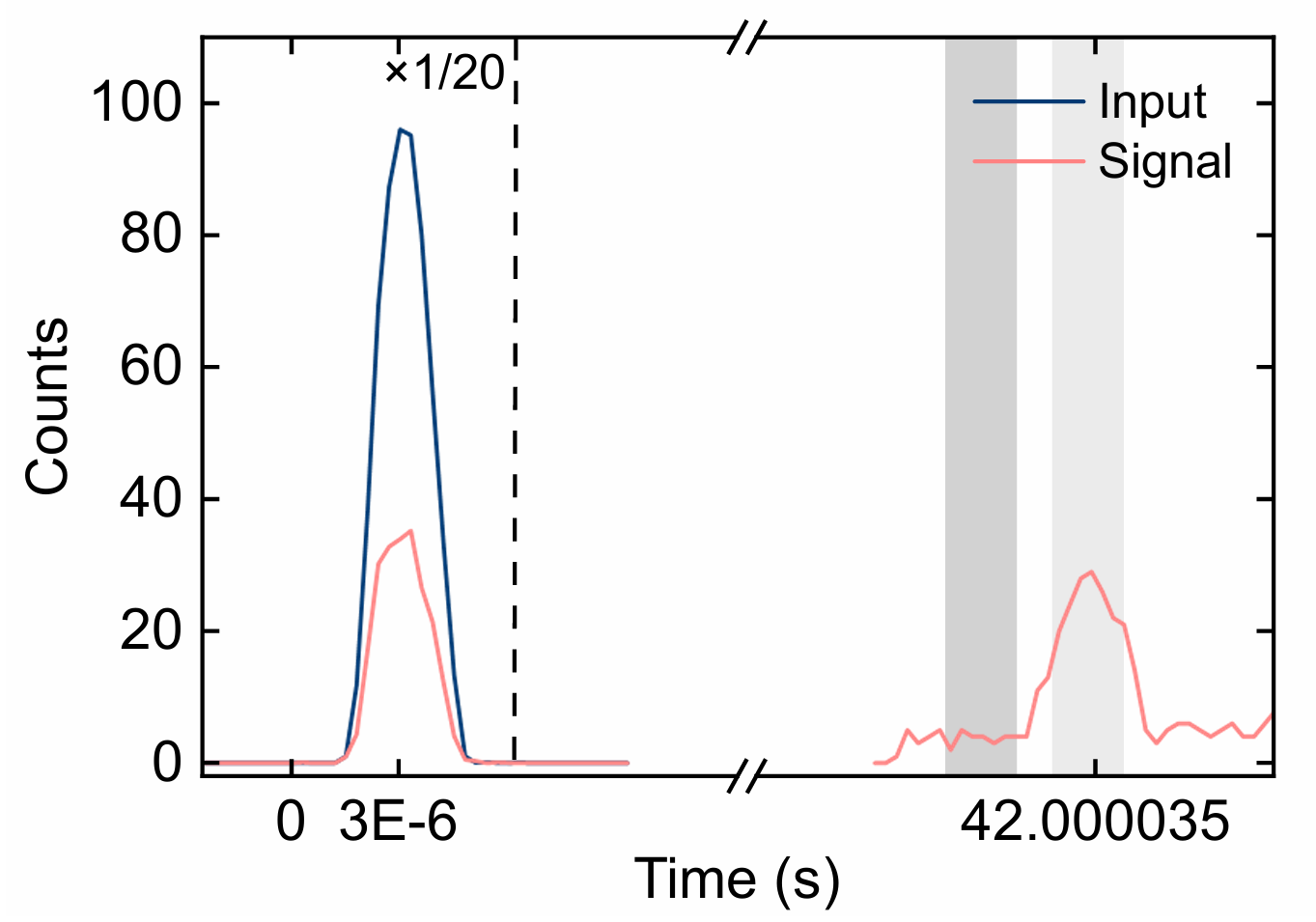}
\end{subfigure}
 
    \caption{Minute-scale storage of single-photon level inputs in a $^{151}$Eu$^{3+}$:Y$_2$SiO$_5$ crystal. (a) Diagram of the experimental setup. Single photons are stored in the memory crystal with a $1/e$ lifetime of 27.6 s, based on a combination of the NLPE protocol, DD, and the ZEFOZ transition. (b) Storage of weak coherent pulses over 42 s. For input pulses with a mean photon number of $\mu = 4.14$, the measured SNR was $5.54 \pm 1.38$. When normalized to one photon per pulse, this gives an SNR of $1.34 \pm 0.33$. Reproduced under the terms of the CC BY 4.0 license~\cite{lv2025minutescalephotonicquantummemory}.}
    \label{fig7}
\end{figure*}

Recently, Lv \textit{et al.} further combined NLPE, DD, and ZEFOZ fields all together, achieving storage of single-photon level inputs on the minute scale using a bulk $^{151}$Eu$^{3+}$:Y$_2$SiO$_5$ crystal \cite{lv2025minutescalephotonicquantummemory}. 
The schematic of the experimental setup is shown in Fig.~\ref{fig7}(a). The relative direction of the memory crystal and the DC magnetic field at 1.3 T is carefully adjusted using two goniometers, with the same ZEFOZ magnetic field as in Ref.~\cite{Zhong2015}. The signal and control beams are overlapped at 1.5$^\circ$ inside the memory crystal, and the retro-reflected signal beam is extracted for fiber coupling. RF fields from surrounding solenoid coils implement DD to protect spin coherence. The retrieved photonic signal undergoes temporal gating based on an acousto-optic modulator (AOM) and two-stage spectral filtering (implemented with a narrow band-pass filter and a filter crystal) before final single-photon detection.

In this work, the DD pulse adopt adiabatic complex-hyperbolic-secant (CHS) pulses \cite{PhysRevA.31.2753,PhysRevA.69.022321} to achieve efficient and uniform manipulation of the entire inhomogeneously broadened ensemble. The CHS pulse is further combined with the universally robust (UR4) sequence \cite{PhysRevLett.118.133202} to achieve extreme robustness against pulse errors. The experimental results show that the $1/e$ storage time reaches $27.6 \pm \SI{0.5}{\second}$, improving the quantum storage time record \cite{Körber2018,PhysRevLett.126.090501} by a factor of more than 60. At a memory time of \SI{5.6}{\second}, the time-bin qubit fidelity reached $88.0\%$, significantly exceeding the classical strategy limit (82.1\%) for $\mu = 1.16$, thereby unambiguously demonstrating the quantum nature of the storage. A single-photon-equivalent SNR of 1.34 is obtained at a storage time of \SI{42}{\second} (Fig.~\ref{fig7}(b)), which could enable a fidelity beyond the classical limit of 2/3~\cite{PhysRevLett.74.1259} assuming a perfect single-photon input. This experiment demonstrates the first minute-scale storage of single-photon-level inputs, enabling potential applications for global-scale quantum networks \cite{3cxd-t3g5,z1vn-11m5} and deep-space quantum experiments \cite{Mol_2023,refId0}.

Long-lived storage has likewise advanced in other platforms. A single atom coupled to a cavity has stored a single-photon qubit with \(22\%\) efficiency and a coherence time exceeding \(100\) ms \cite{Körber2018}, while cold atomic ensembles acting as emissive memories have reached a \(1/e\) storage lifetime of \(458\pm19\)~ms, with an initial retrieval efficiency of \(58\%\) and an atom--photon entanglement fidelity of \(92.3\%\) \cite{PhysRevLett.126.090501}.


\section{Conclusion and Outlook}
\label{sec4}

After decades of sustained effort, REICs, especially Eu$^{3+}$:Y$_2$SiO$_5$ crystals, have demonstrated the longest ensemble spin coherence time (6 hours), the longest light storage time (1 hour), and the longest quantum storage time (27.6 s), positioning them as highly promising candidates for transportable quantum memories.

Further extending the storage lifetime and overall performance will require several advances, including theoretical studies of new storage protocols and decoherence mechanisms, crystal growth with stronger absorption and narrower inhomogeneous broadening, optimized ZEFOZ transitions, and more efficient DD sequences for large ensembles.

The current ZEFOZ transition for Eu$^{3+}$ at 1.3 T remains susceptible to higher-order magnetic-field fluctuations and suffers from significant inhomogeneous broadening across the sample. Future work should seek ZEFOZ transitions with enhanced robustness against external field noise to further extend the Hahn-echo spin $T_2$. Concurrently, the development of better crystal samples with narrower spin homogeneous linewidths would also contribute to longer Hahn-echo spin $T_2$ \cite{10-20240116}.

Although NLPE enables full utilization of the natural absorption for quantum storage, the inherently weak absorption of Eu$^{3+}$—further reduced at high fields by the complete lifting of degeneracy—still limits the absorption and storage efficiency. Impedance-matched optical cavities offer a promising route to efficient storage in weakly absorbing media \cite{PhysRevA.82.022310,PhysRevLett.110.133604}. Recent work on impedance-matched microcavity-enhanced AFC memories has already achieved quantum storage efficiencies beyond 80\% \cite{Meng2026}. Future efforts to combine NLPE with microcavities could thus yield long-lived quantum memories with both high efficiency and long lifetime.

On the practical side, the development of compact cryostats and magnets, along with integrated memories of high device density \cite{ou2026random,z6lc-qw2d}, is essential for real-world applications, particularly for space- and satellite-based deployment. 

Eu$^{3+}$:Y$_2$SiO$_5$ crystals provide a flexible platform for both quantum repeaters and transportable quantum memories. Recently, Zhu \textit{et al.} have also demonstrated a metropolitan-scale multiplexed quantum repeater, ``XingHan 2.0,'' using this material, achieving the violation of Bell's inequality with matter-matter entanglement over a distance of 14.5 km \cite{Zhu2026}. Transportable quantum memories not only offer an alternative channel for quantum communication but also provide viable mobile access to future optical-band quantum networks. This vision points toward a novel architecture for future quantum networks—``\textbf{quantum repeater backbones + transportable storage nodes}''—that closely parallels the classical telecommunications infrastructure of today.

This review has focused on long-lived quantum memories based on rare-earth ion ensembles. However, recent studies of single rare-earth ions have also shown considerable promise for realizing long-lived single-spin qubits \cite{O’Sullivan2025,PRXQuantum.4.010323,Ruskuc2022,ulanowski2026cavityenhanced}, offering a potential alternative route toward long-lived quantum memories featuring deterministic quantum gates.

\section*{Declarations}

\subsection*{Availability of data and material}
All data presented in this review are derived from previously published publicly available studies. Relevant original data can be obtained from the references cited within the manuscript. No new data were generated in this review article.

\subsection*{Competing interests}
The authors declare that they have no competing interests.

\subsection*{Funding}
This work is supported by the Strategic Priority Research Program of the Chinese Academy of Sciences (No. XDB1740000), HFNL Self-Deployed Project (No. ZB2026020200), the Quantum Science and Technology-National Science and Technology Major Project (No. 2021ZD0301200), the National Natural Science Foundation of China (Nos. 12222411).

\subsection*{Authors' contributions}
Y.-Y. S., P.-X. L. and Z.-Q. Z. wrote the manuscript; Z.-Q. Z. and C.-F. L. supervised the project.

\subsection*{Acknowledgements}
Z.-Q.Z acknowledges the support from the Youth Innovation Promotion Association CAS.
\bibliography{sn-bibliography}  

\end{document}